# Exciton Localization in Extended π-electron Systems: Comparison of Linear and Cyclic Structures

Alexander Thiessen[1]*, Dominik Würsch[2], Stefan-S. Jester[3], A. Vikas Aggarwal[3], Alissa Idelson[3], Sebastian Bange[2], Jan Vogelsang[2], Sigurd Höger[3], John M. Lupton[1,2]

[1]Department of Physics and Astronomy, University of Utah, Salt Lake City, UT 84112, USA

[2]Institut für Experimentelle und Angewandte Physik, Universität Regensburg, 93040 Regensburg, Germany

[3]Kekulé-Institut für Organische Chemie und Biochemie der Universität Bonn, 53121 Bonn, Germany

**Corresponding Author**

*alex@physics.utah.edu

**ABSTRACT**

We employ five π-conjugated model materials of different molecular shape – oligomers and cyclic structures – to investigate the extent of exciton self-trapping and torsional motion of the molecular framework following optical excitation. Our studies combine steady-state and transient fluorescence spectroscopy in the ensemble with measurements of polarization anisotropy on single molecules, supported by *Monte Carlo* simulations. The dimer exhibits a significant spectral red-shift within ~100 ps after photoexcitation which is attributed to torsional relaxation. This relaxation mechanism is inhibited in the structurally rigid macrocyclic analogue. However, both systems show a high degree of exciton localization but with very different consequences: while in the macrocycle the exciton localizes randomly on different parts of the ring, scrambling polarization memory, in the dimer, localization leads to a deterministic exciton position with luminescence characteristics of a dipole. *Monte Carlo* simulations allow us to quantify the structural difference between the emitting and absorbing units of the π-conjugated system in terms of disorder parameters.

**TOC GRAPHICS**

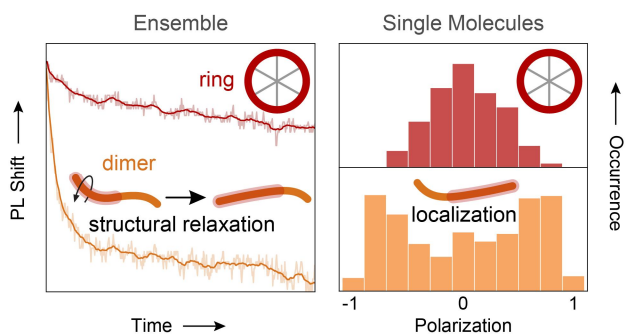

**INTRODUCTION**

Understanding the photophysics of conjugated polymers is crucial to their technological application in organic light-emitting diodes and photovoltaic cells, but is also of interest from a fundamental understanding of light-matter interaction in non-trivial macromolecular geometries.[1-5] Excited state dynamics are inherent to such applications and ultimately determine the efficiency of devices. Some of the most relevant processes after photoexcitation are summarized in Figure 1a. Following excitation (1) by a photon, the excited state electronic density is quickly followed by the nuclei initiating phonon modes. Due to strong electron-phonon coupling in π-conjugated materials the excited state relaxes through the vibrational states manifold and becomes trapped (2) by the rearranged nuclei in form of an exciton.[6] The exciton can undergo intramolecular (3) energy transfer of coherent nature or incoherently in terms of a Förster based hopping mechanism.[7-10] In addition, molecular reorganization, such as planarization or torsional relaxation (4), can further lower the energy of the excited state.[11,12] Both mechanisms constitute a loss of excitation energy and are ideally avoided in functional devices.[13,14] While exciton self-trapping (2) can occur on timescales of <100 fs,[5,15,16] processes (3) and (4) are generally two to three orders of magnitude slower (Figure 1b).[10,12,17,18] The similarity of characteristic timescales makes it challenging to distinguish between the competing processes of energy transfer and torsional relaxation solely on the basis of time-resolved spectroscopies.[19] In principle, it should be possible to distinguish between these three processes with the right choice of model systems and experimental methods of investigation. Here, we study the extent of exciton localization and structural relaxation upon excitation by utilizing π-conjugated model systems of similar building blocks composed in a variety of different geometries, shown in Figure 2. Based on a carbazole-bridged phenylene-ethynylene-butadiynylene repeat unit (synthesis described elsewhere),[20] the monomers are arranged either in linear or in cyclic fashion. In addition, a ring-shaped structure is considered in which the monomers are disconnected and free to rotate.

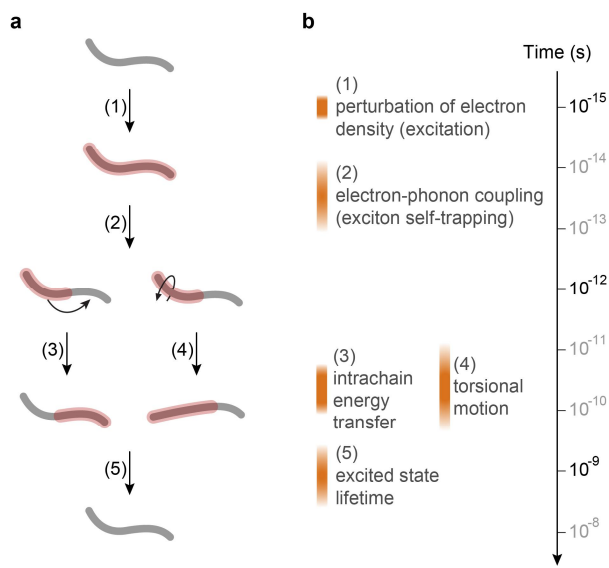

**Figure 1. Overview of possible excitonic processes in π-conjugated macromolecules after photoexcitation and the characteristic timescales of these processes**. **a**, The electric field of a photon perturbs the electron density of the π-system and promotes an electron to a higher energy level (1). Electron-phonon coupling induces localization of the excited state (exciton self-trapping) anywhere on the molecule to a size smaller than the absorbing unit (2). Intramolecular energy transfer propagates the excited state between chromophores within the molecule (3). Torsional motion can lower the energy of the excited state (4). **b**, Typical time scales (orange bars) for the processes shown in panel a.

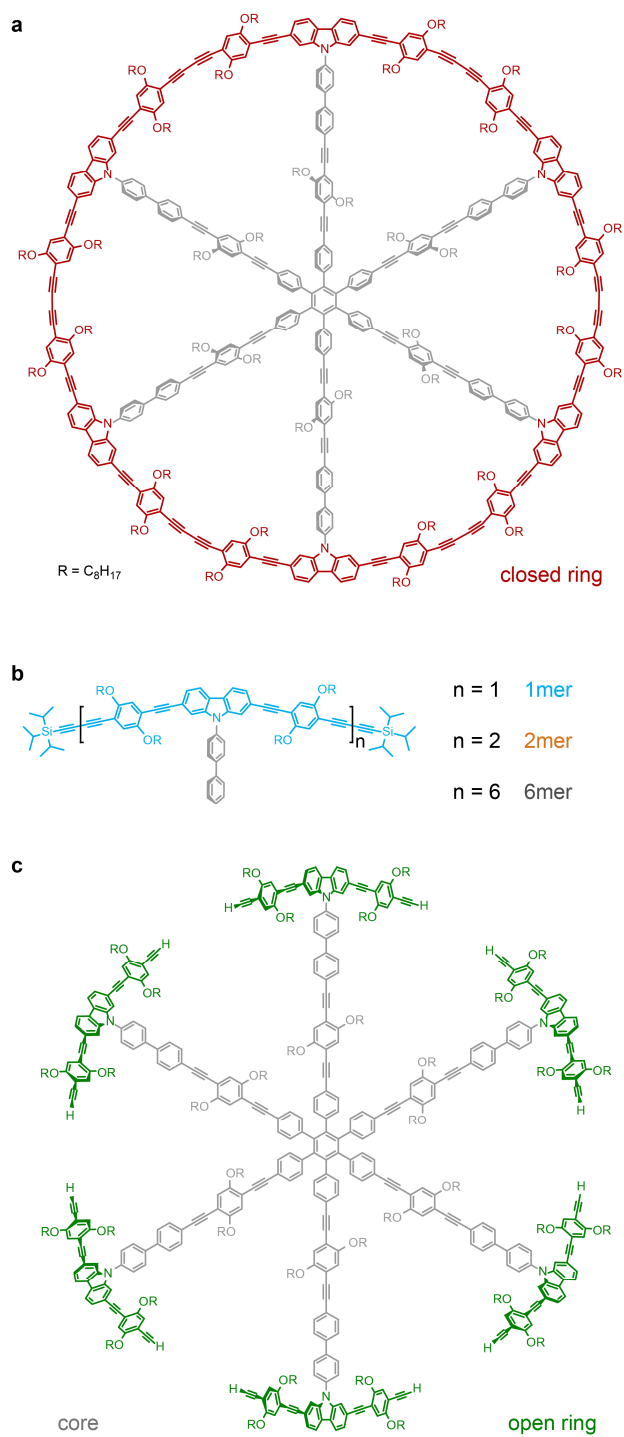

**Figure 2. Chemical structures of the model systems to differentiate between excitonic localization processes in extended π-conjugated macromolecules**. **a**, The closed ring model system is based on a carbazole-bridged phenylene-ethynylene-butadiynylene scaffold (ring rim, red) stabilized by six spokes which emerge from a central hub (gray). The phenyl groups in the

spokes connecting to the carbazole units are rotated out of the ring plane leaving the rim π-conjugation disconnected from the structural centerpiece. **b**, Oligomers constitute the linear model system of the ring's rim. While the monomer corresponds to one sixth of the rim, the hexamer covers the full length of the rim. All oligomers are terminated with 2-(triisopropylsilyl)ethynyl end groups. **c**, The open ring carries monomer units (green) attached to the spokes (gray), whereas the core has only a carbazole unit at the end of each spoke. Individual spoke segments are allowed to rotate around the internal axis.

The interplay between structural rigidity and electronic structure in such large macromolecules is complex and not always intuitive.[16,21-24] A conjugated perfectly circular structure would not be expected to fluoresce from the $S_1$ state since this transition is dipole forbidden.[16,25,26] Such a suppression of fluorescence has indeed been observed in porphyrin-based macrocycles, and in thiophene-based structures.[23,26-28] In contrast, our carbazole ring structures do show strong fluorescence, which we have previously attributed to spontaneous symmetry breaking in the excited state which localizes the molecular exciton on a rim segment, enabling fluorescence.[20] This symmetry breaking corresponds to a breach of the Franck-Condon principle since the transition dipole becomes dependent on nuclear coordinate.[16] An estimate of the magnitude and nature of structural relaxation can be derived from the Stokes shift between absorption and emission, and in particular from the transient fluorescence spectrum. Further insight into the influence of the heterogeneous molecular structure on the emitting and absorbing units can be provided by dissecting the ensemble to the single molecule level.[29] Single molecule polarization techniques can probe the relative orientation of absorbing and emitting units to provide information on exciton localization.[30-32] We demonstrate here that exciton localization on the single molecule level is required to reconcile the polarization anisotropy in excitation with the linear dichroism measured in emission in a statistical analysis. Careful selection of methods and model materials allows us to discern between different excited state relaxation mechanisms arising from structural and torsional relaxation and intramolecular energy transfer.

**MATERIALS AND METHODS**

**Sample Preparation**

All materials were stored and handled in an inert (nitrogen) atmosphere glovebox. Ensemble solutions were prepared by dissolving the model compounds in toluene (EMD Chemicals, OmniSolv) at concentrations of ~$10^{-6}$ – $10^{-7}$ mol/L. Single-molecule samples were prepared by dispersing the analytes in poly(methyl-methacrylate) (PMMA, $M_w$ = 97 kDa, $M_n$ = 46 kDa, PDI 2.1 from Sigma Aldrich Co.) on glass cover slips. The substrates were cleaned in a Hellmanex III (Hellma Analytics) MilliQ water solution (volume fraction 0.02), rinsed with MilliQ water and transferred to a UV-ozone cleaner (Novascan, PSD Pro Series UV) to oxidize residual organic contaminants. The analyte and PMMA were dissolved in toluene at concentrations of ~$10^{-12}$ mol/L and 1 g/L, respectively. The solution was dynamically spin-coated onto the glass cover slips at 2,000 rpm, resulting in film thicknesses of approximately 50 nm and an average analyte density of 40 molecules in an area of 50 × 50 μm$^2$. Single-molecule samples were mounted in a microscope and measured under ambient conditions.

**Wide-Field Polarized Excitation Fluorescence Microscopy**

Wide-field fluorescence microscopy was performed using an inverted microscope (Olympus IX71) with a 1.35 NA oil immersion objective (Olympus, UPLSAPO 60XO). A fiber-coupled diode laser (PicoQuant, LDH-C-405) operating in continuous wave mode at 405 nm was used as the excitation source. Laser fluorescence was removed from the excitation light by a narrow-excitation filter (Semrock Inc., 405/10 nm MaxDiode, LD01-405/10) and a Glan-Thompson polarizer was used to provide linearly polarized excitation light. An electro-optical modulator (FastPulse Technology Inc., 3079-4PW) in combination with a λ/4 waveplate was used to rotate the polarization light.[33] Wide-field illumination over an area of ~80 × 80 μm² at an irradiance of 100 mW/cm² on the sample was achieved by passing the laser beam through the back port of the microscope and a dichroic mirror (AHF analysentechnik AG, RDC 405 nt) focused onto the back focal plane of the objective. Sample emission was collected through the same objective and dichroic mirror, magnified 1.6× and detected with an EMCCD camera (Andor, iXon3 897). Scattered laser light was removed with a fluorescence band-pass filter (Semrock Inc., Edge Basic LP 405 long-pass filter) prior to detection. Diffraction limited spots of single molecules covered ~ 2 × 2 pixels at a resolution of 160 nm/pixel and an overall magnification of 96×. Polarization modulation measurements were performed by rotating the excitation light in the sample plane by 180° over periods of 20 s. EMCCD images were recorded as a function of polarization angle. For the analysis, an area of 5 × 5 pixels for individual spots was integrated from which a local background of a surrounding area of 13 × 13 pixels was removed. Data analysis was performed using custom MATLAB code.[34]

**Linear Dichroism Measurements**

Linear dichroism of single molecules was recorded with the same fluorescence microscope as described above for wide-field excitation and imaging, but instead in confocal arrangement with a circularly polarized and collimated laser beam coupled into the objective. In order to find the analyte molecules, a fluorescence image of 20 × 20 μm² (50 nm/pixel resolution) was recorded by stage scanning (Physik Instrumente, model P-527.3CL) at integration times of 2 ms/pixel. The excitation power of the diffraction limited spot was 50 nW at a wavelength of 405 nm. Spots with homogeneous fluorescence intensity were manually selected and subsequently placed in the laser focus to record emission of single analyte molecules. The emission signal was split into two orthogonal polarization channels with a polarizing beam splitter (Thorlabs, CM1-PBS251). The signal was recorded with avalanche photodiodes (Picoquant, τ-SPAD-20) with a time-correlated single-photon counting module (Picoquant, HydraHarp 400).

**Time-Resolved Fluorescence Spectra and Transient Fluorescence Anisotropy**

Time resolved measurements were performed using a Hamamatsu streak camera system comprising a spectrograph (Bruker Corporation, Bruker 520IS), a streaking unit (Hamamatsu Corporation, C5680) and an ORCA-ER CCD camera (Hamamatsu Corporation, C4742-95). The samples were excited using a frequency-doubled Ti:sapphire femtosecond laser system (APE GmbH, HarmoniXX and Coherent Inc., Chameleon Ultra II), operating at 390 nm (for the monomer and open ring excitation) or 405 nm (for dimer, hexamer and closed ring excitation). Fluorescence anisotropy measurements were performed in an L-format configuration.[35] To account for the polarization-dependent instrument sensitivity an isotropic emission pattern was measured, generated with horizontally polarized excitation. All measurements were carried out on ~$10^{-6} – 10^{-7}$ mol/L solutions in 10 mm quartz cells (Starna Cells Inc., Quartz Fluorometer Cells Micro, 28F-Q-10). The maximal time resolution of the streak camera system was found to be 5.2 ps under excitation in the UV.

**Monte Carlo Simulations of Excitation and Emission Polarization Histograms**

Histograms of polarization anisotropy in excitation (modulation depth) and linear dichroism in emission were simulated by constructing a set of 50,000 randomly distributed monomer or dimer molecules, respectively. A single monomer molecule was constructed from ten equally long linear dipole segments; a single dimer molecule consisted of two monomer units (20 total dipole segments). Continuous bending was induced with the angle $\theta_t$ between the first and the last dipole segment of one monomer unit. Two monomer units forming a dimer were allowed to have a random dihedral angle $0 \leq \varphi \leq 2\pi$. Structural disorder was introduced by assuming a normal distribution of bending angles with a mean bending angle $\mu$ and a standard deviation $\sigma$ for a given set of molecules. Molecules were randomly oriented in three-dimensional space. From a set of 50,000 monomer or dimer molecules, 17,000 monomers and 8,000 dimers with orientations along the sample plane were selected by imposing a maximum out of plane angle of 20° (average dipole vector angle). This assumption is based on the experimental observation and simulations of the molecules being largely oriented flat on the sample. Anisotropy in excitation was calculated from the dipole vector components interacting with the electric field vectors rotating in the two-dimensional $x,y$-sample-plane. The $z$-component of the electric fields was assumed to be negligible owing to the fact that excitation polarization modulation depth was measured in wide-field excitation with no $z$-polarization component present. Signal fluctuations were introduced from a Poisson-distributed signal and background with experimentally obtained mean values. Mean bending angles and standard deviations of a given set were changed in steps of $\{\Delta\mu, \Delta\sigma\} = 1°$, and the calculated $M$ histogram was compared to the experimental $M$ histogram using a least-squares test. In contrast to excitation anisotropy, linear dichroism was measured under confocal geometry (with an objective numerical aperture of 1.35). Histograms of linear dichroism in emission were thus generated from the simulation data by mapping the dipole orientation from the sample plane onto the objective's back plane, appropriately averaging over ray positions in the back plane. Two perpendicular polarization directions ($x,y$) in the objective's back focal plane were used to calculate the linear dichroism $D = (x - y)/(x + y)$ for individual molecules.

**RESULTS AND DISCUSSION**

**Model Systems and Room Temperature Steady State Ensemble Spectroscopy**

Our model systems are based on a carbazole-bridged phenylene-ethynylene-butadiynylene scaffold[20,36] assembled into effectively one-dimensional (oligomers), two-dimensional (closed ring) and three-dimensional (open ring) geometries as shown in Figure 2. The design of the macrocycle molecule (closed ring) was chosen primarily to provide shape persistency to the π-conjugation of the closed ring's rim, Figure 2a. Six spokes emerging from a central hub manifold are connected to the *N*-phenyl-carbazole bridged phenylene-ethynylene-butadiynylene elements. Because the phenyl groups of the spokes adjacent to the *N*-carbazole units are rotated out of the molecule plane, π-conjugation of the rim is effectively disconnected from the center moiety. The perimeter of the closed ring is translated into a one-dimensional configuration in the three different oligomer structures shown in Figure 2b. The smallest unit (monomer) composed of one sixth of the closed ring's rim with the *N*-phenyl-carbazole unit at its center. The longest unit (hexamer) is equal to the full perimeter of the closed ring. All oligomers are terminated with 2-(triisopropylsilyl)ethynyl end groups. The open ring molecule, shown in Figure 2c, constitutes the precursor of the closed ring with the rim dissected at six points between the spokes. Each spoke carries a monomer building block with the full structure resembling six anchors attached to a common centerpiece. This arrangement becomes particularly interesting because each anchor is anticipated to rotate around the shank axis thus limiting the anchor's crown to a plane perpendicular to the spokes plane. Finally, we also considered the model system of the structural centerpiece, the core of the macrocycle. The core contains only the carbazole unit and not the phenylene-ethynylene building block of the monomer.

The results of room-temperature absorption and emission spectroscopy of the compounds in toluene solution are summarized in Figure 3. The emission spectra of the oligomers show a bathochromic shift from the monomer to the dimer suggestive of improved electronic delocalization. However, there is almost no spectral shift between the dimer and the hexamer, indicating that excitonic coherence[37] does not extend significantly beyond the size of the dimer. At the same time the dimer exhibits the strongest Stokes shift (~40 nm; 240 meV) of all oligomers. The substantial blue-shift of ~20 nm (120 meV) of the open-ring emission compared to the monomer is attributed to the absence of two ethynyl units from the terminal 2-(triisopropylsilyl)ethynyl protecting groups in the open ring, resulting in stronger confinement. The exact values of absorption and emission peak positions are summarized in Table 1. Interestingly, the spectra of the spokes and the open ring structure are remarkably similar. We show below that the emission in the open ring structure indeed originates from the carbazole anchor crowns and not from the spokes.

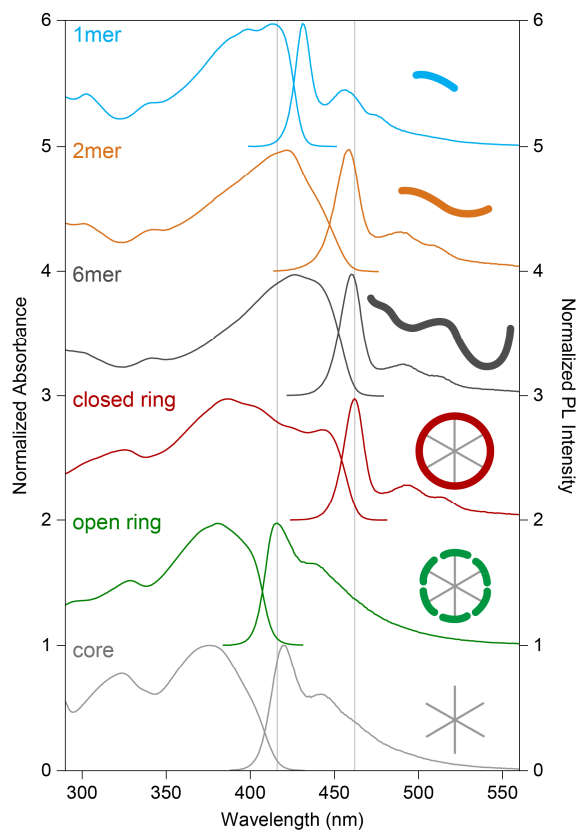

**Figure 3. Steady-state optical spectra of the model systems**. The increasing size of the chromophoric system causes a progressive red shift of the low-energy absorption peak from monomer to dimer over hexamer to closed ring. In contrast, the red shift of the emission saturates with chain length for the dimer, suggesting that the emissive units in the hexamer and closed ring structures are effectively of the same size as the emissive unit in the dimer. The open-ring emission peak is slightly blue shifted compared to the monomer due to the lack of 2-(triisopropylsilyl)ethynyl end groups in the open ring. Although absorption and emission of the core and the open ring structure appear to be very similar, consideration of the polarization anisotropy shows that emission in the open ring occurs from the crown (the monomer unit) and not from the spokes. Gray lines are added as a visual aid.

**Table 1.** Absorption and photoluminescence (PL) peak wavelength, PL lifetimes, PL quantum yields (PLQY) and radiative lifetimes measured in toluene solutions at room temperature.

|  | Absorption maximum (nm) | PL maximum (nm) | PL lifetime (ns) | PLQY (%) | Radiative lifetime (ns) |
|---|---|---|---|---|---|
| 1mer | 413 | 431 | 0.97 | 84 | 1.15 ± 0.07 |
| 2mer | 422 | 459 | 0.58 | 69 | 0.84 ± 0.06 |
| 6mer | 426 | 460 | 0.47 | 66 | 0.71 ± 0.06 |
| closed ring | 443 | 462 | 0.60 | 71 | 0.85 ± 0.06 |
| open ring | 381 | 414 | 0.94 | 62 | 1.52 ± 0.12 |
| core | 376 | 419 | 1.03 | 75 | 1.37 ± 0.09 |
| error | ± 2 | ± 2 | ± 0.01 | ± 5 |  |

**Time-Resolved Fluorescence Spectra and Fluorescence Depolarization**

While the emission peak energy only changes between monomer and dimer and then remains constant for the larger structures (Figure 3), the absorption peak of the carbazole compounds experiences a continuous red-shift in the series going from monomer over dimer and hexamer to the ring (Table 1). The substantial Stokes shift of the dimer between absorption and emission indicates a high degree of structural relaxation[38]. Insight into the relaxation process can be provided by time-resolved fluorescence spectroscopy as presented in Figure 4. The dimer experiences a strong transient red-shift of the emission peak following photoexcitation, readily visible in the streak camera images (white arrow in panel a). The relative peak energy $\Delta E(t) = E(t) - E_{t=0}$ is plotted in Figure 4b. It is not immediately obvious why the strongest energy shift occurs in the dimer, totaling over 40 meV, whereas the transient Stokes shift of the monomer or ring amounts to no more than 4 meV. Naïvely, one might anticipate the hexamer to exhibit a more pronounced spectral shift, owing to the larger and more disordered molecular structure as well as the possibility of on-chain energy transfer. It is likely that the red shift in the dimer is associated with torsional motion between monomer units that cannot occur in the monomer and is slowed down in the hexamer due to the increased molecular size. Similarly slow structural relaxation on the timescale of >100 ps has been reported for trifluorene,[18] conjugated porphyrin oligomers[11] and polyfluorene[17] previously, although torsional relaxation is usually found to be somewhat faster in π-conjugated oligomers.[39-41]

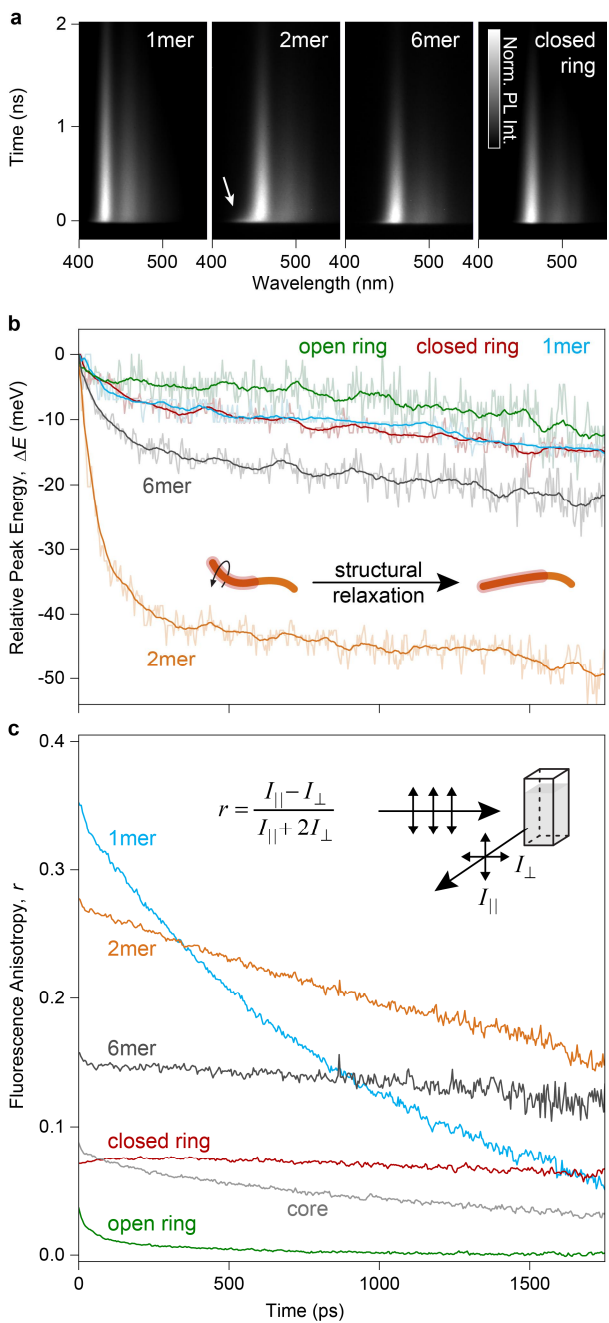

**Figure 4. Time-resolved ensemble PL spectra and fluorescence polarization anisotropy transients in dilute solution**. **a**, Streak camera images of the monomer, dimer, hexamer and closed ring molecules obtained in toluene solution, at room temperature, plotting the PL intensity versus wavelength and time. The dimer shows a pronounced red shift of the PL maximum (white arrow) which is caused by structural (most likely torsional) relaxation. This shift is not observed

in the hexamer or the ring structures. **b**, Temporal shift of the emission peak plotted as the relative energy change $\Delta E(t) = E(t) - E_{t=0}$. The dimer shows the largest energy loss during the first 200 ps suggesting a high level of structural relaxation. The spectral shift of the core emission is indistinguishable from that of the closed ring and is therefore not shown. **c**, The fluorescence anisotropy $r$, defined and measured as shown in the inset, quantifies the polarization memory in the excited state. The initial values of $r$ depend on intramolecular depolarization that occurs on the fs-timescale which is linked to the geometry of the system. Within the measured time range, the anisotropy decay is governed by rotational diffusion of the molecules in solution which is reduced for the larger systems. A fast depolarization component in the open ring structure due to either rotation of the crown units or energy transfer between them is visible during the first 200 ps. The solutions were excited at 390 nm (for the monomer and open ring) and at 405 nm (for the dimer, hexamer and closed ring).

Although the steady state photoluminescence (PL) peak energy is virtually the same for the dimer, hexamer and the closed ring, further conjugation increases the pool of electrons contributing to the exciton wavefunction. This increase raises the emission oscillator strength[37] and hence lowers the radiative lifetime (Table 1) from $\tau_{1mer} = 1.15$ ns over $\tau_{2mer} = 0.84$ ns to $\tau_{6mer} = 0.71$ ns. Despite the structural similarity of the monomer and the open ring's crowns, the radiative lifetime of the latter $\tau_{open} = 1.52$ ns is longer than $\tau_{1mer}$, whereas the lifetime of the closed ring $\tau_{closed} = 0.85$ ns is almost identical to that of the dimer.

It is interesting to look at the dynamics of the orientation of the emitting transition dipole of the model systems in order to identify processes occurring immediately after photoexcitation, in particular interchromophoric coupling and intramolecular energy transfer. Figure 4c provides an overview of the temporal fluorescence anisotropy decay of the compounds in solution. An isotropic distribution of linear dipoles is expected to show an anisotropy $r = (I_\parallel - I_\perp)/(I_\parallel + 2I_\perp)$ of $r = 0.4$.[42] Here, $I_\parallel$ and $I_\perp$ indicate vertical and horizontal polarization of the luminescence under vertically-polarized laser excitation. The maximum value of 0.4 can only be reached if the absorption and emission dipoles are colinear. When this is not the case, the anisotropy deviates according to[35]

$$r = \frac{3\cos^2\beta - 1}{5}$$

where $\beta$ is the angle between the absorption and emission transition dipole moments. Accordingly, $r$ can have values ranging from 0.4 to −0.2 corresponding to $\beta$ values of 0° and 90°, respectively, where $r = 0$ implies complete depolarization. The monomer shows an initial anisotropy value of $r_{t=0} = 0.35$ in Figure 4c. The subsequent decrease of $r$ with time arises due to rotational diffusion of the molecules in the solvent. As chain length increases to the dimer and hexamer, a smaller initial $r$-value is determined due to greater bending of the molecular object, followed by a *slower* subsequent decay with time due to the larger radius of gyration. The closed ring displays no discernible dynamics in anisotropy around $r = 0.1$. Such a value is expected for ultrafast depolarization of fluorescence in a planar system randomly distributed in space,[22,26,43] implying delocalization of excitation energy on a timescale < 5 ps (below the time resolution of our system), which can be caused by the mechanisms discussed in the introduction including internal conversion. A similar behavior is observed for the core (gray curve). However, for this compound an additional decay below $r = 0.1$ is observed, which likely originates from the smaller effective radius of gyration in solution and the reduced structural rigidity compared to the ring. If the emission in the open ring were to occur from the spokes, the fluorescence anisotropy decay would show very similar behavior to that observed for the core. The open ring structure, in contrast, exhibits complete fluorescence depolarization to $r \sim 0$ within 200 ps. This complete depolarization implies rotational displacement of the chromophores out of the molecular plane, adding a further degree of freedom for fluorescence depolarization by energy transfer.

Both transient PL spectra as well as time resolved polarization anisotropy reveal substantial differences between the different model systems which are masked when simply comparing steady-state ensemble PL spectra. The measured initial anisotropy values (limited by our time resolution of ~5 ps) provide crucial information on the polarization memory following processes occurring on the femtosecond to picosecond time scales which can include exciton self-trapping, intrachain energy transfer and structural relaxation.[9,10,44] However, much more can be learned by directly studying the absorbing and the emitting units of individual molecules which are inherently interrelated with the molecular shape.

**Single Molecule Polarization Spectroscopy**

Ensemble averaging effects can be overcome by studying morphology related characteristics of π-conjugated systems on the single molecule level.[29,45,46] Dispersing the molecules in poly(methyl-methacrylate) at picomolar concentrations provides an amorphous environment in an inert polymer matrix and sufficient intermolecular spacing so that single molecules can be examined optically. The molecular concentration in the ~50 nm thick matrix film was adjusted to yield well separated, diffraction limited emission spots. We employ excitation polarization modulation and emission polarization spectroscopy to study characteristics of the absorbing and emitting units, respectively. As sketched in Figure 5a, the absorption polarization ellipse of a single molecule can be traced by modulating its fluorescence intensity $I$ with excitation polarization following

$$I(\alpha) \propto 1 + M \cos[2(\alpha - \phi)],$$

where $\alpha$ is the polarization angle in the plane of excitation, $\phi$ is a phase factor and $M$ the modulation depth. This definition of $M$ is equivalent to taking the difference and sum ratio of maximal and minimal emission intensities, as stated in Figure 5a. Intensity modulation traces of individual molecules give a value of $0 \leq M \leq 1$ with $M = 1$ signifying a linear absorption dipole and $M = 0$ an unpolarized absorber.[47,48] These values are plotted in a histogram for the different compounds in Figure 5b-f, with the number of single molecules measured stated in each panel. As expected, the monomer exhibits near-perfect linearly polarized absorption ($M \rightarrow 1$). As the chain length increases to the dimer, the chromophores become more bent, lowering $M$ (Figure 5c). The core exhibits a very broad histogram, with both high and low anisotropy values present. For both the closed ring and the open ring, the intensity modulation with polarization angle approaches zero. Non-zero values in the histogram may conceivably arise due to slight out-of-plane orientation of the molecules and the finite signal-to-background ratio in the single-molecule measurement. However, none of the closed ring molecules show high $M$ values which would be expected for rings lying perpendicular to the sample plane.[49] We therefore conclude that out-of-plane orientations of the closed rings do not arise, possibly due to shear forces acting on the ring molecules during spin coating or due to effective phase separation between these large molecules and the host matrix, i.e. molecules migrating to the matrix surface during solvent evaporation. In contrast, the emissive units in the open ring are flexible compared to the closed ring. Rotation of the chromophores of the open ring out of the molecular plane, as discussed in Figure 4c, will introduce anisotropy yet at the same time lower intensity since the transition dipoles along the rim units point out of plane, decreasing the effective absorption cross-section. To test this hypothesis, panel d superimposes the distributions for the brightest (white) and darkest (black) 50 % of the open ring molecules on top of the average histogram. Indeed, the brighter the molecules, the more the $M$ values tend to zero. Similarly to our conclusion from the fluorescence anisotropy decays in solution (Figure 4c), this observation implies that it is indeed the rim unit in the open ring which is optically active and not the spokes, since the spokes themselves cannot easily rotate out of the molecular plane. It is interesting to compare the open ring and the spokes directly since, in their unperturbed state, they have virtually the same $D_{6h}$ symmetry. The scatter to high anisotropy for the spokes structure compared to the open ring must therefore be a result of reduced structural integrity leading to the formation of less isotropic molecular objects.

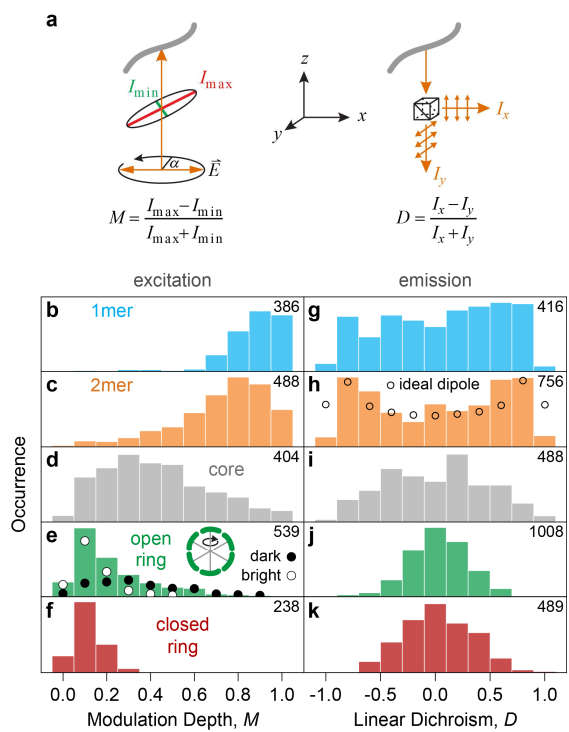

**Figure 5. Geometries of absorbing and emitting units probed on the single-molecule level**. **a**, Sketch of the excitation polarization ellipse of individual molecules, which is probed by rotating the polarization plane of the excitation light and recording the PL modulation depth $M$ (left). The orientation of the emitting dipoles is determined by splitting the single-molecule PL into two orthogonal polarization channels and is characterized by the linear dichroism $D$ (right). **b-f**, Histograms of excitation polarization anisotropy for the different model systems. A clear trend towards lower $M$ values is apparent when going from the monomer, a linearly polarized absorber, to the fully conjugated ring, which acts as an almost perfectly isotropic absorber. **e**, Splitting the $M$ distribution of the open-ring structures into two equally large subsets of bright and dark molecules reveals the cause for the broadening of the histogram to higher $M$ values. Molecules with anchor crowns rotated out of the ring plane appear as less isotropic, having higher modulation depths, but also couple radiation less effectively to the far field, thereby appearing dimmer. **g-k**, Following the increasingly isotropic shape of the molecules from the monomer to the closed ring, a peak of linear dichroism around a value of 0 emerges in the single-molecule histograms. Whereas perfectly isotropic systems always result in $D = 0$, the histogram of an ideal dipole (open circles in panel **h**) peaks at values close to $D = \pm 1$. Only the dimer approaches this case (**h**). The number of molecules in a histogram is given in each panel. All samples were excited at 405 nm.

The molecular morphology of the model systems is clearly reflected in the excitation polarization modulation depth histograms. A similar trend is observed for the *emitting* units classified by measuring the linear dichroism $D$ under circularly polarized excitation as defined in Figure 5a. Here, the PL is separated into two orthogonal polarization components by a polarizing beam splitter. Depending on the relative orientation between emitter and detectors, and the overall degree of linearity in the emitter, the linear dichroism can assume values ranging from $-1$ to $+1$. A histogram of randomly distributed ideal dipoles would peak at values of $D = \pm 1$ with a pronounced dip around $D = 0$.[50] This ideal distribution, arrived at by a straightforward *Monte Carlo* simulation, is superimposed in the measured histogram in Figure 5h as open circles. In contrast, in an isotropic system with no preferential emission polarization, linear dichroism will always be zero. Given the signal to noise level in the experiment, the monomer and dimer $D$ distributions in Figure 5g,h are fairly close to expectation for a single linear transition dipole. The core structure, which constitutes a multichromophoric emitter, is only weakly polarized (panel i). $D$ histograms for the ring structures in Figure 5,j,k tend to peak around $D = 0$ indicating isotropic emission polarization characteristics. The fundamental difference between the fluorescence anisotropy ($r \sim 0.1$) in solution and the linear dichroism in single-molecule films peaking at $D = 0$ as observed for the closed ring structure results from the fact that in solution, the two-dimensional molecular objects are randomly distributed, whereas in the thin-film experiments they appear to lie preferentially in the *x-y* plane with little projection of the transition dipole into the *z*-direction. Interestingly, the linear dichroism histogram for the open rings (panel j) is virtually identical to that for the closed rings (panel k). We have shown previously that individual closed ring molecules act as single photon sources, yet no correlation between excitation and emission polarization is observed.[20] The closed-ring exciton localizes dynamically and non-deterministically to the size of the dimer exciton, resulting in random emission polarization from anywhere on the perimeter. Similar behavior was recently reported in photosynthetic complexes and bent *β*-phase polyfluorene chains.[32,51] The deviation of $D$ values from zero in the closed ring structure has been shown to originate from photoinduced symmetry breaking, leading to a certain degree of deterministic localization of the emissive exciton.[20]

Despite the fact that the monomer structure is less bent than the dimer, it is not the monomer but rather the dimer that approaches the expected *D* distribution of an ideal dipole, with the *D* histogram peaking around ± 1. Most notably, comparison of the monomer's and dimer's *M* and *D* histograms appears to suggest a discrepancy between the geometry of the absorbing and the emitting units of the molecules. While in absorption the monomer behaves more like an ideal dipole, showing a trend to overall higher polarization anisotropy values and a smaller scatter (Figure 5b) than the dimer (Figure 5c), in emission, the situation is reversed: panel h more closely resembles the simulated ideal case (open circles) than does panel g. In order to assess the degree of structural modification upon excitation, we carried out *Monte Carlo* simulations of the *M* and *D* histograms for the monomer and dimer.

**Simulation of Modulation Depth and Linear Dichroism Histograms**

Measurement of excitation polarization anisotropy (modulation depth *M*) directly probes the shape of the absorbing chromophores. In contrast, the emissive species may be prone to structural localization effects (exciton self-trapping)[6] or energy transfer[9]. Considering the large size of our model molecules, occurrence of geometric (structural) relaxation in the PMMA film is highly unlikely.[52] We model the effect of excited-state relaxation in the molecules as outlined in Figure 6a. A monomer unit is discretized numerically and defined as being made up of ten equally long, linear dipole segments. Two such units represent a dimer molecule, which are allowed to have a random dihedral angle $\varphi$ between them. This assumption is based on the fact that a series of multiple triple bonds effectively flattens the angle-dependent rotation barrier allowing for virtually any orientation.[53-55]

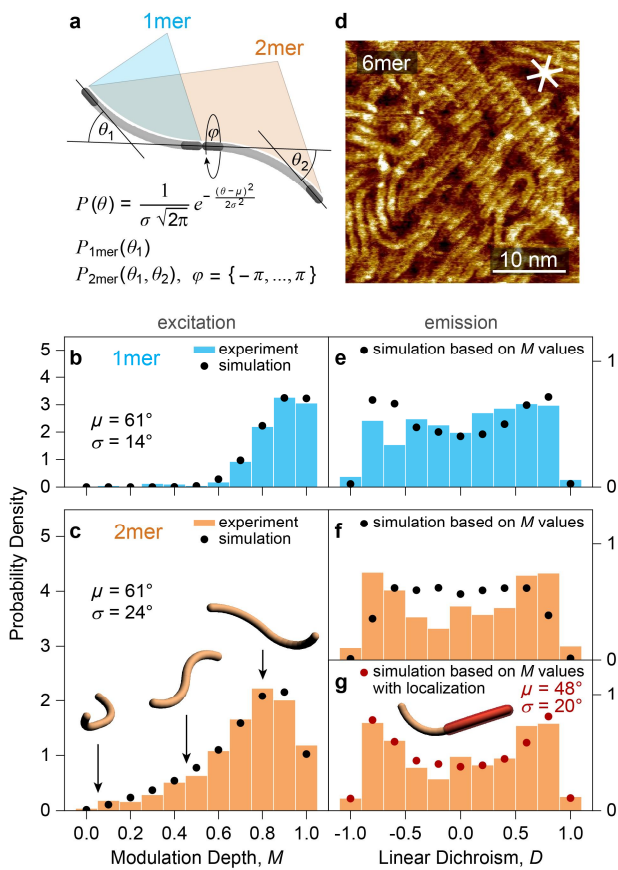

**Figure 6. *Monte Carlo* simulations of the single-molecule excitation and emission anisotropy histograms of monomer and dimer, showing the effect of exciton localization in the excited state**. **a**, A monomer unit is arbitrarily broken down into ten equally long linear dipole segments with an overall bending angle $\theta_1$ between the first and the last segment. The dimer consists of two such units with independent individual bending angles $\theta_1$ and $\theta_2$, and a random dihedral angle $\varphi$ between the two units. Assuming a normal distribution of bending angles $\theta \sim \mathcal{N}(\mu, \sigma^2)$ with mean angle $\mu$ and standard deviation $\sigma$ leaves only two free simulation parameters which are adjusted to fit the experimental distributions. **b**,**c**, Comparison of experimental data (bars) and simulation results (dots). The simulated distributions arise from 17,000 and 8,000 randomly

chosen monomer and dimer conformations, respectively. The best agreement between experiment and simulation was achieved with the same mean bending angle of 61° for both monomer and dimer but with significantly different standard deviations. The inset shows three representative dimer conformations from the simulation. **d**, STM image of the hexamer on a graphite substrate showing that longer molecules can undergo significant bending and distortion, despite the attractive interaction with the crystalline substrate (white lines indicate the highly ordered pyrolytic graphite main axis direction). **e**,**f**, Linear dichroism experimental data (bars) and simulation results (dots). The same simulated data set as in excitation anisotropy (panels b,c) is considered in the computation of linear dichroism. The experimental *D* histogram of the dimer is not well reproduced by the molecular geometries extracted from the fit to excitation anisotropy *M* (panel c), suggesting that the emissive region of the dimer is less bent than the absorbing unit. **g**, Assuming exciton localization on the less distorted unit of the two monomers making up the dimer, the measured distribution can be reproduced closely by the simulation (red dots). An example of such a localization process is illustrated in the inset on a molecular conformation employed in the simulation. Here, the emissive exciton is localized to the less distorted unit of the dimer, marked in red.

Variation of the molecular shape is introduced by a bending angle $\theta_1$ within the monomer unit, defined as the angle between the first and the last discretized element of one unit. For simplicity we assume continuous bending of the segments belonging to one repeat unit.[57] Disorder is introduced into the simulation through a normal distribution of bending angles $\theta_1 \sim \mathcal{N}(\mu_{1mer}, \sigma^2_{1mer})$ leaving only two variable parameters in the simulation when comparing to the measured distribution of $M$ values: the mean bending angle $\mu$ and the standard deviation $\sigma$. The dimer has two independent bending angles $\theta_1$ and $\theta_2$ for each of the two repeat units but both are restricted by the same normal distribution $\{\theta_1, \theta_2\} \sim \mathcal{N}(\mu_{2mer}, \sigma_{2mer}^2)$. Finally, each simulation set contains 17,000 monomer and 8,000 dimer molecules, respectively, interacting with the electromagnetic field polarized in the sample plane. The molecules are randomly oriented in the sample plane with a maximum average out of plane angle of 20°, which provides the best agreement of experimental and simulation results. The best fit parameters are identified by minimizing a least-squares test of simulation results performed in steps of $\{\Delta\mu, \Delta\sigma\} = 1°$.

For clarity, the results of the simulations are plotted in Figure 6 as dots on top of the experimental histograms of Figure 5, using identical bin sizes. Our simulation reproduces the modulation depth ($M$) histograms remarkably well, as displayed in Figure 6b,c. In general, we find that increasing the bending angle $\theta$ shifts the peak of the histogram to lower $M$ values equally for monomer and dimer. However for identical monomer and dimer bending angles, in the dimer, an increase in frequency of low anisotropy values and a decrease in frequency around $M = 1$ occurs due to the increased molecular complexity and the resulting structural disorder. This complexity comes from the presence of two monomer units in the dimer instead of one and the random dihedral angle $\varphi$ between them. Fits to the monomer and dimer modulation depth distributions yield the same mean bending angles $\mu_{1mer} = 61°$ and $\mu_{2mer} = 61°$, respectively, as would be expected for identical molecular building blocks. However, the width of the dimer histogram is best reproduced with larger standard deviation of the bending angles for the dimer compared to the monomer ($\sigma_{2mer} = 24°$ vs. $\sigma_{1mer} = 14°$), indicating that the dimer constitutes a more disordered system.

Three representative dimer conformations taken from the ensemble of simulated molecules, are sketched in Figure 6c. While these dimers appear to be strongly distorted, we note that similarly bent shapes are actually observed in scanning tunneling microscopy (STM) of the hexamer, shown in Figure 6d. This image was acquired on highly-oriented pyrolytic graphite substrate; see ref.[20] for experimental details. Smaller molecules such as the dimer have a strong tendency for crystallization during the self-assembly on the substrate,[20] so that such structural disorder cannot be observed for shorter chains in STM. However, we argue that it is plausible that the dimer also experiences such bending, similar to that seen here in the hexamer, when embedded in an amorphous polymer matrix as used in the single-molecule experiments.

Using the set of molecules from the $M$ histogram simulations we can compute the linear dichroism in emission, $D$, as shown in Figure 6e,f for the monomer and dimer, respectively (dots). Without making any assumptions regarding exciton localization, i.e. assuming that the emitting region of the molecule is identical to the absorbing region, we note that the *measured D* distribution of the monomer (panel e) is much better reproduced by the simulation than the *measured D* distribution of the dimer (panel f), even though the fits of the simulations to the $M$ (excitation) histograms (panels b, c) are of comparable quality. In particular, the simulated $D$ values (dots in panel f) do not reproduce the measured dip around $D = 0$ for the dimer. This deviation of the simulation of $D$ values from the measurement for the dimer provides evidence for a pronounced difference in the geometry of absorbing and emitting units in the dimers.

We can account for the effect of excited-state localization by assuming that emission occurs from the least-deformed region of the dimer as sketched in red in Figure 6g. This simple restriction drastically changes the shape of the simulated $D$ histogram (red dots) calculated from the same set of (simulated) molecular conformations used in the excitation polarization anisotropy ($M$) histogram in panel c. The mean bending angle of the region of the dimer on which exciton localization occurs ($\mu_{loc} = 48°$, $\sigma_{loc} = 20°$) is substantially smaller than the mean bending angle of the monomer $\mu_{1mer}$ (61°). Our analysis follows simple geometrical arguments without taking into account molecular dynamics in the excited state[6]. Without considering such structural changes following excitation, our representation of localization is likely to underestimate the magnitude of spatial localization, but nevertheless offers a quantifiable estimate in terms of disorder parameters $\mu$ (overall average bending) and $\sigma$ (scatter of bending between molecules). The simulations indicate that in the ground state, monomer units in the dimer are bent to the same degree as in the monomer. Following photoexcitation and the resulting structural reorganization, a significant difference arises: the emitting units of the dimer are bent much less than in the monomer. We note that the fact that we do not observe any change in ensemble polarization anisotropy of the dimer (Figure 4c) on a timescale comparable to that of the dynamic red shift (Figure 4b) suggests that the structural relaxation in the excited state does not significantly contribute to a change of emission dipole orientation. The excited-state localization which gives rise to the marked difference between the single-molecule excitation and emission polarization anisotropy histograms must therefore occur on sub-picosecond timescales and most likely arises from intramolecular redistribution of excitation energy.[6]

**CONCLUSIONS**

We have investigated the degree of structural relaxation and exciton self-trapping in $\pi$-conjugated model systems of different molecular shape. Steady-state emission spectra from solutions of the dimer, hexamer and the closed ring indicate that the extent of the emissive unit is the same in all three systems. However, the emission peak of the dimer experiences a significant spectral red-shift of ~40 meV within ~100 ps after excitation, which we attribute to torsional relaxation. This relaxation pathway is inhibited in the structurally rigid ring and is also absent in the hexamer. We propose that such torsional relaxation in the hexamer is significantly impeded due to the larger size of the molecule. Single molecule polarization spectroscopy in excitation and emission allows an estimate of the degree of exciton localization in the excited state without having to resort to complete molecular dynamics simulations.[6,58] While the closed ring behaves almost like an isotropic emitter because the exciton localizes dynamically on different parts of the ring,[20] localization in the dimer leads to an emissive unit behaving almost as an ideal linear dipole. *Monte Carlo* simulations of excitation and absorption anisotropy histograms for the monomer and the dimer suggest that chromophores in both molecules are bent to the same degree ($\mu_{1mer} = \mu_{2mer} = 61°$), but are more disordered in the dimer ($\sigma_{2mer} = 24°$) than in the monomer ($\sigma_{1mer} = 14°$). However, after exciton localization the emissive unit in the dimer is on average less bent ($\mu_{loc} = 48°$, $\sigma_{loc} = 20°$) than in the monomer, where exciton delocalization is limited by the smaller molecule size. Our custom-made π-conjugated model systems allow us to differentiate between self-trapping and torsional relaxation involving motion of larger parts of the molecule. Torsional relaxation occurs on timescales of hundreds of picoseconds and does not appear to affect the orientation of the transition dipole moment. Instead, the marked difference between ground and excited-state transition dipole orientation on the single-molecule level is likely due to intramolecular redistribution of excitation energy with little structural rearrangement. The comparison of ensemble and sub-ensemble spectroscopy techniques in combination with microscopic control of rigidity in model systems demonstrated here provides a general route for future investigations of the photophysics of conjugated polymers.

**Corresponding Author**

*alex@physics.utah.edu

**Author Contributions**

The manuscript was written through contributions of all authors. All authors have given approval to the final version of the manuscript.

ACKNOWLEDGMENT

We are indebted to the Volkswagen Foundation for funding of collaborative materials development, and thank the German Science Foundation for support through the Research Training Group 1626 and through the SFB 813. JML gratefully acknowledges support from the European Research Council through Starting Grant MolMesOn No. 305020.